\documentstyle[PASJadd,epsf]{PASJ95}
\begin{document}
\setcounter{page}{1}

%
%

\title{Evolution of Clusters of Galaxies:
Mass Stripping from Galaxies and Growth of Common Halos}

\author{Tomohiro {\sc Sensui},$^1$ Yoko {\sc Funato},$^2$
 and Junichiro {\sc Makino}$^1$\\[1ex]
$^1$ {\it Department of Astronomy, The University of Tokyo, Bunkyo-ku, Tokyo 113-0033}\\
{\it E-mail(TS): sensui@grape.c.u-tokyo.ac.jp}\\
$^2$ {\it General Systems Sciences, Graduate Division of International and Interdisciplinary Studies,}\\{\it The University of Tokyo, Meguro-ku, Tokyo 153-8902}}

%
%

\abst{%
We investigated the evolution of clusters of galaxies using self-consistent $N$-body simulations in which each galaxy was modeled by many particles.
We carried out simulations for about 20 cases using different initial conditions.
In all simulations, clusters were initially in virial equilibrium.
We found that more than half of the total mass escaped from individual galaxies within a few crossing times of the cluster, and that a diffuse halo was formed.
The growth rate of the common halo depended on the size of individual galaxies only weakly.
The stripping of the mass from galaxies was mainly due to the interaction of galaxies, not due to the effect of the tidal field of the cluster potential.
The amount of stripped mass was larger for galaxies in the central region than for those in the outer region, since the interactions were more frequent in the central region.
As a result, a positive correlation between the distance from the center and the mass of the galaxy developed.
The volume-density profile of the common halo is expressed as $\rho\propto r^{-1}$ in the central region.
This mass distribution is consistent with the mass distribution in clusters estimated using X-ray observations.}

%
%

\kword{%
Galaxies: clusters: general ---
Galaxies: evolution ---
Galaxies: interactions}

\maketitle
\thispagestyle{headings}

%
%

\section{Introduction}

In this paper, we present the result of a series of $N$-body simulations of the evolution of clusters of galaxies.
We investigated the evolution of clusters of galaxies in which each galaxy was modeled by many particles.

The formation of clusters of galaxies have been studied by numerous researchers both by $N$-body simulations and combined SPH (Smoothed Particle Hydrodynamics)+$N$-body simulations or hydrodynamics+$N$-body simulations (see Bertschinger 1998 for a recent review).
These simulations typically follow the nonlinear growth of the density fluctuation under a given cosmology.
The majority of these simulations have regarded a cluster as an essentially ``smooth'' system, composed of $N$-body and SPH particles.
This is simply because the resolution in mass and/or space in many of these simulations is insufficient to resolve individual galaxies.

As a result, the internal degree of freedom of individual galaxies is ignored.
However, whether such a ``smooth'' approximation is acceptable or not is questionable, since typical clusters contain only $10^{2}$--$10^{3}$ galaxies.
In addition, if we are to understand the evolution of individual galaxies in clusters, we have to resolve individual galaxies.
It has been long known that the morphology of galaxies shows a correlation with the local number density of galaxies (Dressler 1980).
This could be due to the different initial condition (Evrard et al.\ 1990), but the interactions with nearby galaxies and tidal field of the parent cluster certainly play important roles in determining the present-day morphology of cluster galaxies.

Funato et al.\ (1993, hereafter FME) performed simulations of clusters of galaxies in which each galaxy was expressed using a fairly large number of particles.
They found that individual galaxies lose a large fraction of their masses in a few crossing times of the cluster.
The mass which escaped from galaxies formed a cluster-wide common halo.
Once this halo had been formed, galaxies started to interact with this halo, mainly through dynamical friction.
Thus, the evolution of a cluster of galaxies is very different from that of a cluster of point-mass particles.
In their simulations, mergers were rare and no cD was formed.
This is essentially because the velocity dispersion of the cluster is large.

Bode et al.\ (1994) also conducted several simulations using the mass spectrum of galaxies, and found that a cD-like galaxy formed in the center of the cluster.
Garijo et al.\ (1997) performed similar simulations as those in FME.
Their main interest was in the formation of cD galaxies.
They tried various initial conditions, including cold initial conditions.
In most cases they found that a central cD evolved on a rather short timescale.
The difference of their results can be qualitatively understood as being due to the difference of the initial conditions.

Unfortunately, it is difficult to draw a general picture of the dynamical evolution of a cluster from these studies.
The cases which they chose were too diverse to extract any general evolutional scenario.
In other words, no systematic study of the possible parameter space has been performed.
This is mainly because of the limitation in the available computing resources.
Important parameters include the number of galaxies in the cluster, the ratio between the velocity dispersion of galaxies and that of the cluster, galaxy mass function.

In this paper, we present the results of a systematic study of the evolution of a virialized cluster, as the first step to obtain a quantitative understanding of the evolution of the structure of a cluster of galaxies.
We investigated the structure of the common halo, and the timescale and mechanism of its growth.

In section 2 we describe the initial conditions for cluster models, and in section 3 we describe the numerical method used in our simulation.
In section 4 we present the results.
A summary and discussions are given in section 5.

\section{Initial Conditions}

 \subsection{Overview}

We performed a series of $N$-body simulations on the evolution of clusters which were initially in dynamical equilibrium.
The top-left panel in figure~1 shows an example of the initial conditions.

In all, runs with one exception (run E2), we set the initial number of galaxies in the clusters to be the same.
In addition, all galaxies in a cluster were initially identical.
We systematically changed the ratio between the size of the cluster and that of its member galaxy and compared the results.
The galaxy and cluster models are summarized in table~1.

In runs A, the cluster models were generated from different random seeds.
In other words, all models in runs A are the same, except for the random seed.
However, since a cluster contains only 128 galaxies, not only the positions and velocities of each galaxies, but also the total binding energy and virial radius of the clusters are different.
We performed these runs to see run-to-run variations due to different random seeds.

In order to distinguish the effect of the difference in the distribution of galaxies and that of the size of the cluster, we carried out two additional series of runs (B and C).~
In runs B, we scaled one cluster model to three models which had different radii to see only the effect of the size of the cluster.
In runs C, on the other hand, we scaled three cluster models, which were generated from different random seeds and had different radii, to the same virial radius to see only the effect of the initial discrete distribution of galaxies.

In runs D, we changed the number of particles comprised in one initial galaxy to investigate the effect of two-body relaxation.
In runs E, we set up a common halo in addition to galaxies in the initial cluster to investigate how the evolution of the cluster changes when a massive common halo initially exists.

In the following, we describe the system of units we used (subsection 2.2) and the details of the initial models (subsection 2.3).

 \subsection{Units}

We used a system of units in which $m=G=1$ and $e = -1/4$, where $G$ is the gravitational constant, and $m$ and $e$ are the mass and energy of one galaxy at the beginning of simulations, except for run E1 ($m=0.5$).
In other words, individual galaxies were expressed in the Heggie units (Heggie, Mathieu 1986).
If we assume that the mass unit corresponds to $10^{12}M_{\odot}$ and the length unit to 30 kpc, the time unit corresponds to 110 Myr.

 \subsection{Initial Models of Galaxies and Clusters}

%
%

\begin{table*}[t]\small
 \begin{center}
 Table 1. Initial conditions.
 \end{center}
 \begin{center}
 \begin{tabular}{lcccclrl}
  \hline\hline
  \multicolumn{1}{c}{ID} & \multicolumn{5}{c}{cluster} & galaxy & \multicolumn{1}{c}{note} \\
  \hline
   & $M$ & $R_{\rm vr}/r_{\rm vr}$ & $E_{0}$ & $T_{\rm cr}$ & seed & $n_{\rm p}$ & \\
  \hline
  A1  & 128 & 25.3 & $-$162.2 & 31.7 & a1  & 512 & \\
  A2  & 128 & 23.5 & $-$174.4 & 28.5 & a2  & 512 & \\
  A3  & 128 & 22.5 & $-$182.0 & 26.7 & a3  & 512 & \\
  A4  & 128 & 21.3 & $-$192.2 & 24.6 & a4  & 512 & \\
  A5  & 128 & 21.0 & $-$194.6 & 24.1 & a5  & 512 & \\
  A6  & 128 & 21.0 & $-$194.8 & 24.1 & a6  & 512 & \\
  A7  & 128 & 20.8 & $-$196.6 & 23.8 & a7  & 512 & \\
  A8  & 128 & 20.1 & $-$203.6 & 22.6 & a8  & 512 & \\
  A9  & 128 & 19.9 & $-$206.2 & 22.1 & a9  & 512 & \\
  A10 & 128 & 18.8 & $-$217.8 & 20.4 & a10 & 512 & \\
  A11 & 128 & 18.1 & $-$226.2 & 19.2 & a11 & 512 & \\
  A12 & 128 & 17.0 & $-$240.7 & 17.6 & a12 & 512 & \\
  A13 & 128 & 16.6 & $-$246.1 & 17.0 & a13 & 512 & \\
  A14 & 128 & 16.4 & $-$250.1 & 16.6 & a14 & 512 & \\
      &     &      &          &      &     &     & \\
  B1  & 128 & 25.6 & $-$160.0 & 32.4 & b1  & 512 & \\
  B2  & 128 & 20.0 & $-$204.8 & 22.4 & b1  & 512 & \\
  B3  & 128 & 15.8 & $-$260.0 & 15.6 & b1  & 512 & \\
      &     &      &          &      &     &     & \\
  C1  & 128 & 20.0 & $-$204.8 & 22.4 & c1  & 512 & A1 scaled. \\
  C2  & 128 & 20.0 & $-$204.8 & 22.4 & c2  & 512 & A8 scaled. \\
  C3  & 128 & 20.0 & $-$204.8 & 22.4 & c3  & 512 & A14 scaled. \\
      &     &      &          &      &     &     & \\
  D1  & 128 & 20.8 & $-$196.6 & 23.8 & d1  & 512 & same as A7. \\
  D2  & 128 & 20.8 & $-$196.6 & 23.8 & d1  & 1024 & \\
  D3  & 128 & 20.8 & $-$196.6 & 23.8 & d1  & 4096 & \\
      &     &      &          &      &     &      & \\
  E1  & 128 & 20.0 & $-$204.8 & 22.4 & e1  & 1024 & $m_{\rm gx}=0.5$, \\
      &     &      &          &      &     &      & with a common halo. \\
  E2  & 128 & 20.0 & $-$204.8 & 22.4 & e2  & 1024 & $N_{\rm gx}=64$, \\
      &     &      &          &      &     &      & with a common halo. \\
  \hline
 \end{tabular}
 \end{center}
\end{table*}

  \subsubsection{Galaxy model}

We used a Plummer model as the initial model for galaxies.
Its mass-density profile is given by
\begin{equation}
\rho(r) =
 \frac{3m}{4\pi r_{0}^{3}}
 {\left[1+\left(\frac{r}{r_{0}}\right)^{2}\right]^{-5/2}},
\end{equation}
where $r_{0} = 0.6$ in our units.
For this model, the half-mass radius $r_{\rm h} $ is $0.73$, the virial radius $r_{\rm vr}$ is $1$ and velocity dispersion $\sigma_{\rm gx}$ is $1/\sqrt{2}$.
For the run E1, we set the mass of one galaxy to 0.5, and scaled the velocity dispersion so that the galaxy would satisfy the relation $m\propto\sigma^{1/4}$, i.e., $\sigma_{\rm gx,E1}=(0.5)^{1/4}\sigma_{\rm gx}$.

A galaxy comprises $n_{\rm p}$ particles, where $n_{\rm p}$ is 512 for runs A, B, and C, and 1024 for runs E.~
For runs D, $n_{\rm p}$ is 512, 1024, and 4096 particles for runs D1, D2, and D3, respectively.
Since the initial mass of a galaxy $m$ is 1, except for E1 (0.5), the mass of a particle is 1/512 for runs A, B, C, and D1, 1/1024 for runs D2 and E2, and 1/4096 for the run D3.
In the run E1, the mass of a particle is 1/2048.

  \subsubsection{Cluster models}

For the model of a cluster of galaxies we also used a Plummer model.
For all runs, except for E, the cluster comprised 128 galaxies, and all of the mass was initially attached to the galaxies.
For runs E, half of the total mass was initially attached to the galaxies, and the remaining half formed a common halo.
The common halo also had the Plummer profile with the radius being the same as that of the cluster.
The number of galaxies was 128 for run E1 and 64 for run E2.
As discussed in subsubsection 2.3.1, the mass of galaxies in run E1 was 0.5, and that in run E2 was 1.
Therefore, the mass of a particle in these runs was 1/1024 and 1/2048, respectively.
The total mass $M$ of a cluster was, therefore, 128 in all runs.

We changed the radii of the cluster models as shown in table~1.
The virial radius of the cluster $R_{\rm vr}$ was changed from about $15\, r_{\rm vr}$ to about $25\, r_{\rm vr}$.
The velocity dispersion $\sigma_{\rm cl}$ was scaled so that the cluster model would be in the virial equilibrium state.

The cluster models for runs A1--A14 is random realizations of the same Plummer model with $R_{\rm vr} = 20$.
If the number of galaxies is infinite, all these models should have the same $R_{\rm vr}$.
The actual values of $R_{\rm vr}$ shown in table~1 vary because of small-number statistics.

For the Plummer model with $R_{\rm vr}=20$, the velocity dispersion $\sigma_{\rm cl}$, the total energy of the cluster $E$ and the crossing time $T_{\rm cr}$ are:
\begin{eqnarray}
\sigma_{\rm cl}
 &=& 4/\sqrt{5} = 1.789,\\
E
 &=& \frac{1}{2}M{\sigma_{\rm cl}}^{2}-\frac{M^{2}}{2R_{\rm vr}} = -204.8,\\
T_{\rm cr}
 &=& \frac{2R_{\rm vr}}{\sigma_{\rm{cl}}}
   = GM^{5/2}|2E|^{-3/2} = 10\sqrt{5}.
\end{eqnarray}
Under the same scaling as used for galaxies, the total mass of the cluster corresponds to $1.28\times10^{14}M_{\odot}$.
The velocity dispersion of galaxies in the cluster is $\sim$ 520 km\,sec$^{-1}$, the radius of the cluster is 0.6 Mpc and the crossing time is 2.5 Gyr.

The cluster models for runs B1--B3 are the same random realization of the Plummer model scaled to the radius $R_{\rm vr}=25.6\,r_{\rm vr}$, $20\,r_{\rm vr},$ and $15.8\,r_{\rm vr}$, respectively.
The cluster models for runs C1--C3 were scaled so that they would satisfy $R_{\rm vr}=20\,r_{\rm vr}$.
The original random realizations are those used in runs A (see table~1).
The cluster model for runs D1--D3 was the same as that for run A7.
For runs E1 and E2, $R_{\rm vr}$ was scaled to 20.0.
A summary of the initial conditions is given in table~1.

We followed the evolution of these models in isolation.
In other words, we neglected the possible mass infall from intercluster space.
This treatment is justified for the following reason.
The time scale of dynamical evolution is longer in the outer region of the cluster than that in the inner region.
Therefore, the mass infall affects only the outer region of the cluster.
Unless we look at outermost regions, we can neglect the infall.

\section{Numerical Methods}

 \subsection{Time Integration}

The equation of motion of each particle is
\begin{equation}
 \frac{d^{2}\mbox{\boldmath$x$}_{i}}{dt^{2}}
  = -G\sum_{j\neq i}m_{j}
    \frac{\mbox{\boldmath$x$}_{i}-\mbox{\boldmath$x$}_{j}}
         {(|\mbox{\boldmath$x$}_{i}-\mbox{\boldmath$x$}_{j}|^{2}+\epsilon^{2})^{3/2}}\ ,
\end{equation}
where $\mbox{\boldmath$x$}_{i}$ is the position of the $i$-th particle, $m_{i}$ the mass, and $\epsilon$ the softening parameter.

In our simulations, the total number of particles $N$ was 128$n_{\rm p}$, i.e., 65536, 131072, and 524288 for $n_{\rm p}=512$, 1024 and 4096, respectively, for runs A through D.~
For runs E, $N=262144$ (E1) and 131072 (E2) (see table~1).
The softening parameter $\epsilon$ was 0.025.

For all simulations, we used GRAPE-4 (Makino et al.\ 1997).
For runs with $N=65536$ (A, B, C, and D1) and 131072 (D2), we used the direct summation method.
The time step $\Delta t$ was 1/80.
One time step took about 4 seconds for $N=65536$ and 9 seconds for $N=131072$.

For runs with $N\ge131072$, we used the Barnes--Hut tree algorithm (Barnes, Hut 1986; Makino 1991; Athanassoula et al.\ 1998) to save calculation time.
The opening angle $\theta$ was 0.75, and the time step $\Delta t$ was 1/128.
We also used the tree algorithm for $N=131072$ simulation (D2) to compare the direct method and the tree algorithm.
One time step took about 6 seconds in $N=131072$ and 24 seconds in $N=524288$.

In all runs, the time integration was carried out using the leap-frog method.
The errors in the total energy $\Delta E$ were less than $10^{-4}$ in all runs.
The total energy was conserved very well.
Our simulations were performed with sufficiently high accuracy.

 \subsection{Galaxy Identification}

To see how the galaxies evolve and lose their masses, we need to determine which particles belong to which galaxy.
In other words, we need to determine which particles have escaped.
We calculated the number of escapers by the following procedure.
Initially, each particle belongs to one unique galaxy.
At each time step, we calculated the binding energy of each particle in its parent galaxy.
If the binding energy was positive, we regarded that particle as having escaped from the galaxy.
The binding energy was calculated using those particles that were still bound to the parent galaxy.
Thus, we needed to iterate 5--6 times to stabilize the membership.
Using this algorithm, we could trace the identity of a galaxy, even after 90\% of the mass had escaped.
The limitation of the present method is that we cannot deal with the exchange of particles between galaxies or merging events.
Since the fraction of mass exchanged is expected to be negligible, neglecting the exchange does not affect our result.

We have found several merging events.
We regarded two galaxies as being merged if their distance remained small for one cluster crossing time.
For galaxies regarded as merged, we determined the mass of the merger remnant by applying the above procedure for particles from both galaxies at the same time.
We found that the merging events were rare (a few pairs out of 128 galaxies).

%
%

\begin{figure*}[htbp]
~\hfill \epsfbox{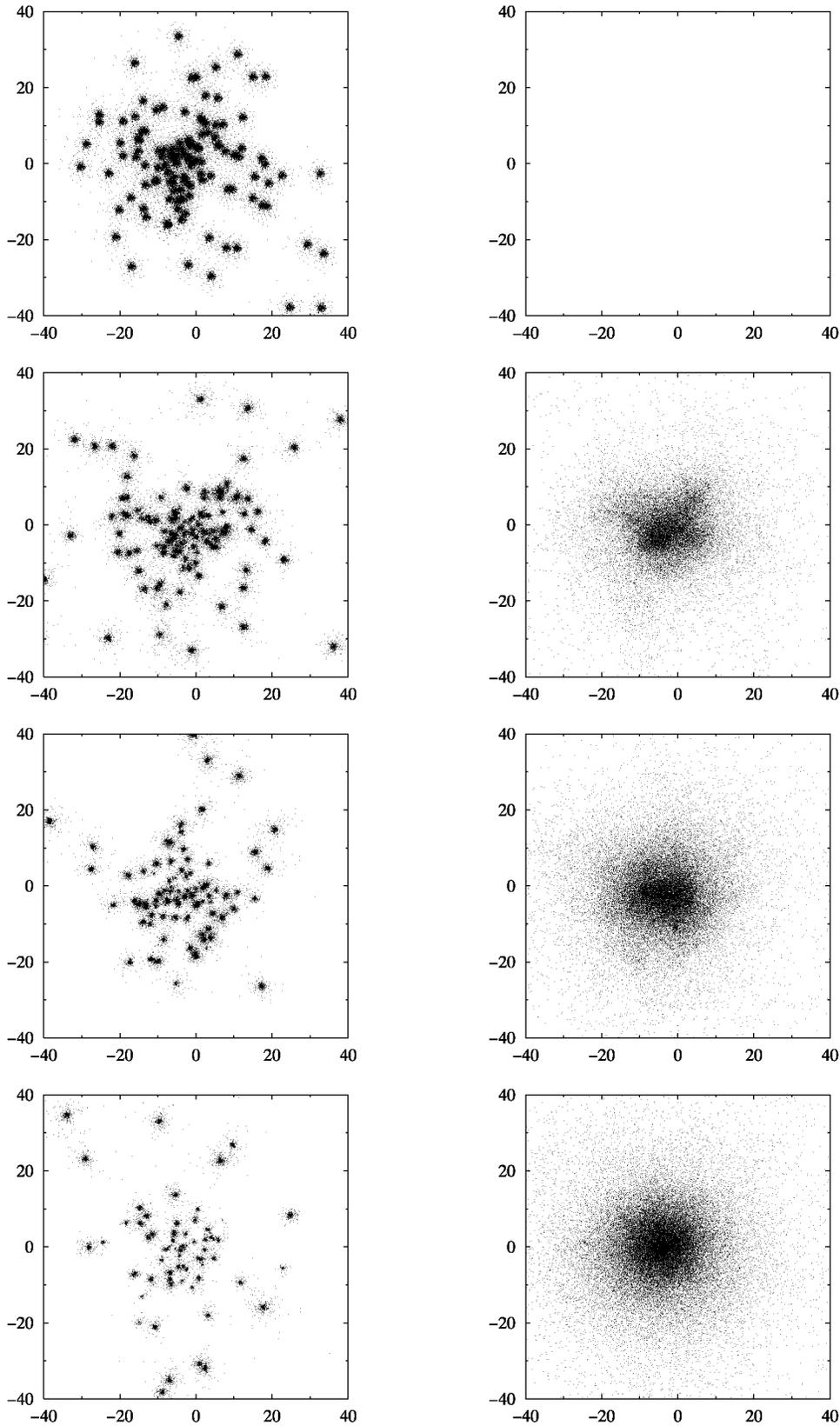} \hfill~
 \caption{%
Snapshots of run A8.
All particles are projected onto the $x$--$y$ plane.
The left-hand side panels show the particles bound to individual galaxies, and right-hand side panels show those escaped to intracluster space.
Each row corresponds to $t=0$ (top), 50, 100, and 200 (bottom).}
\end{figure*}

\section{Results}

 \subsection{Snapshots}

Figure~1 shows snapshots from run A8.
All particles are projected onto the $x$--$y$ plane.
The panels in the left-hand side show particles which were bound to its parent galaxies, and those in the right-hand side show particles which escaped from galaxies to the intracluster space.
We can see that a common halo develops as particles escape from their parent galaxies.
The growth timescale is of the order of the crossing time of the cluster, which is a few Gyrs.

In the following sections, we investigate the structure and growth timescale of the common halo in more detail.

 \subsection{Properties of Clusters}

  \subsubsection{Density profile}

%
%

\begin{figure}[t]
 (a)\\\epsfxsize\columnwidth \epsfbox{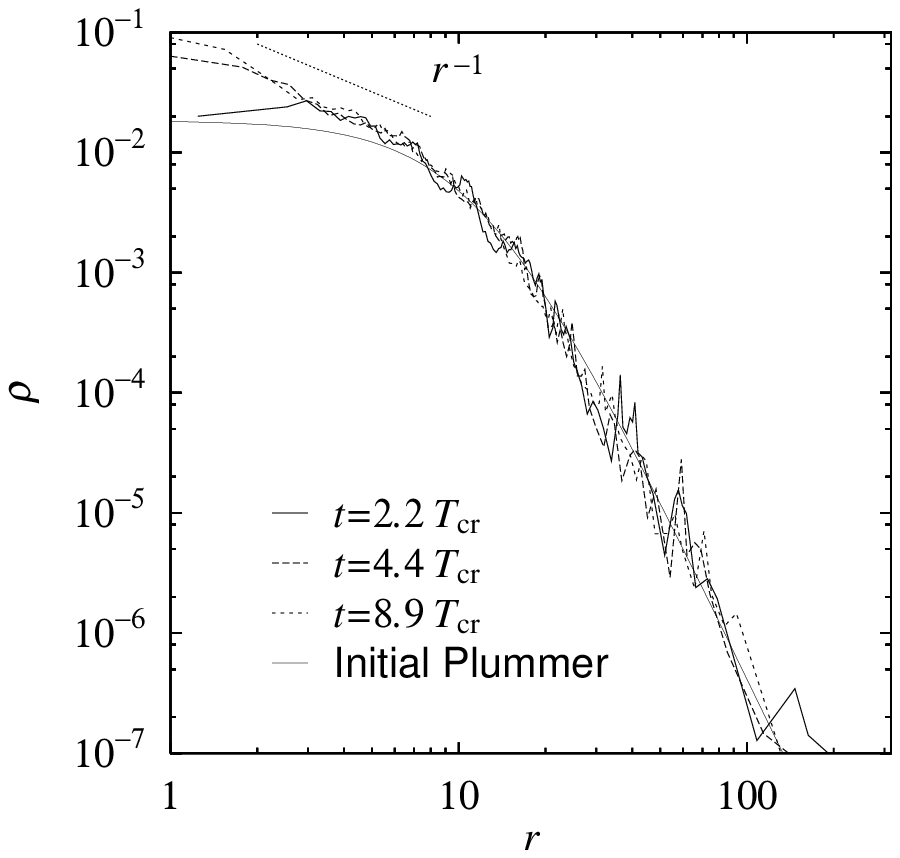}\\
 (b)\\\epsfxsize\columnwidth \epsfbox{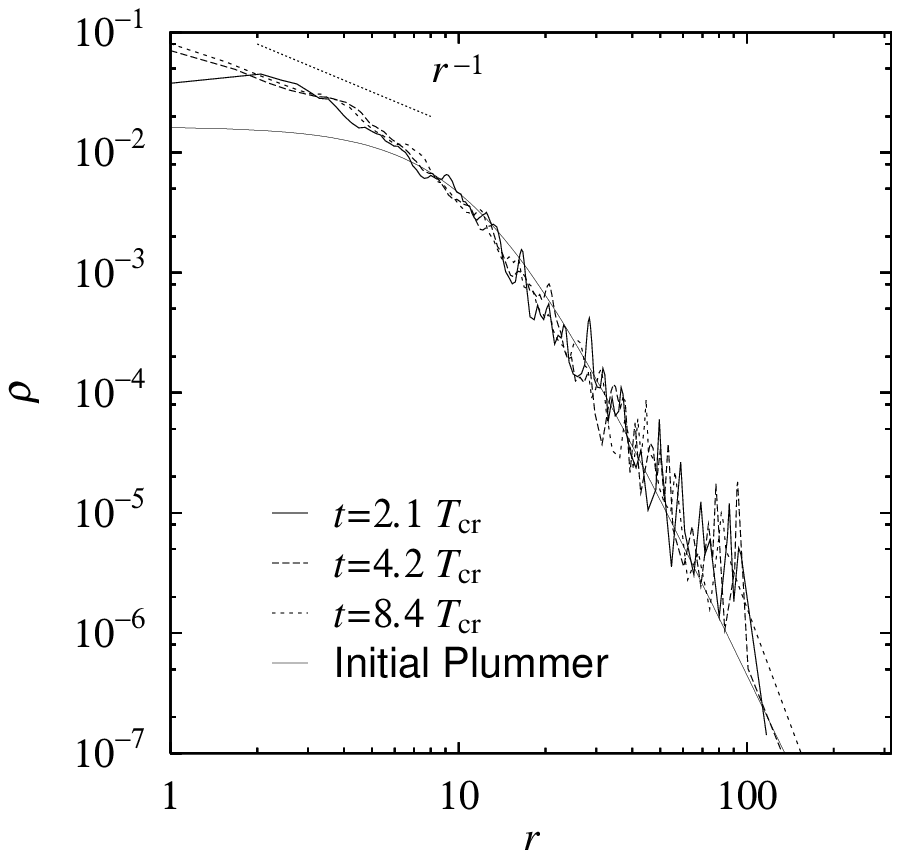}
 \caption{%
Density profiles of clusters for (a) run A8 and (b) run D3.
The solid, long-dashed and short-dashed curves denote the density profiles at $t=50$, 100, and 200, respectively.
Thin solid curve denotes the initial Plummer model.}
\end{figure}

Figure~2 shows density profiles of the clusters for several runs.
The thin solid curve shows the profile of the initial Plummer model, and the rest correspond to $t=50$, 100, and 200.

The slope of the density profile approaches $\rho\propto r^{-1}$ in the inner region, while that in the outer region is almost unchanged from the initial $\rho\propto r^{-5}$ profile.

Real clusters are not isolated and there will be mass infalls, which would make the outer slope less steep (see , e.g., Navarro et al.\ 1996).
As discussed in subsection 2.3, this difference in the outer halo does not affect the evolution of the inner region of the cluster.

The $\rho\propto r^{-1}$ profile of the inner region is remarkable.
It extends almost to the half-mass radius of the cluster, and was formed in all models, including runs B, where we changed the relative size of the galaxies to the cluster.
Figure~3a shows the density profiles of several runs.
We scaled the vertical and horizontal axes so that the initial virial radius would be the same for all runs.
The agreement is extremely good.

The $\rho\propto r^{-1}$ profile also formed in runs E, which had initial common halos with flat cores.
As shown in figure~3b, their density profiles after about 10 $T_{\rm cr}$ are very similar to other profiles which are simulated with no initial common halos.

%
%

\begin{figure}[t]
 (a)\\\epsfxsize\columnwidth \epsfbox{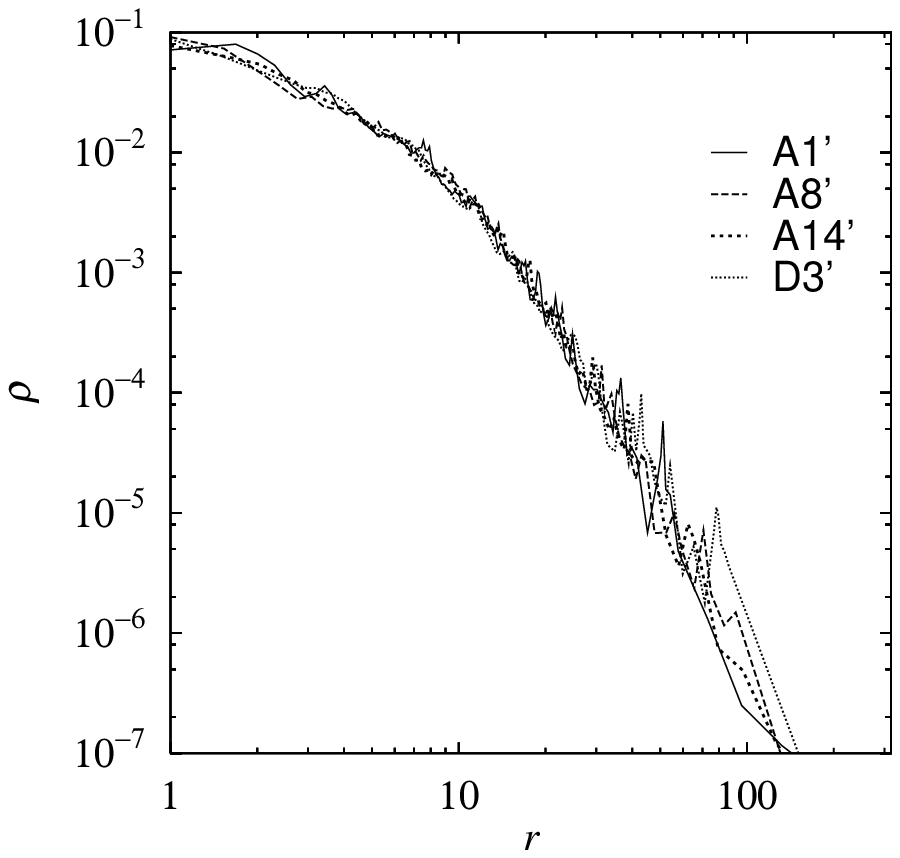}\\
 (b)\\\epsfxsize\columnwidth \epsfbox{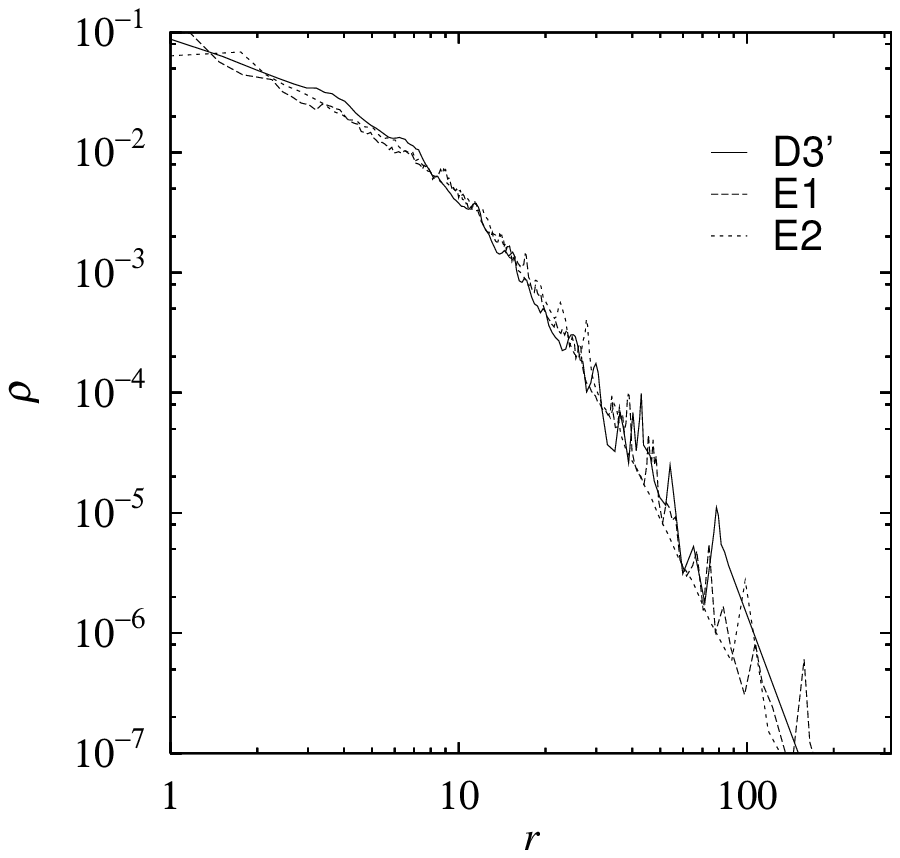}
 \caption{%
Comparison of scaled density profiles at $t=200$ for (a) runs A1 (solid), A8(long-dashed), A14(short-dashed), D3(dotted), and (b) runs D3 (solid), E1 (long-dashed), E2 (short-dashed).}
\end{figure}

Navarro et al.\ claimed that the ``universal'' profile,
\begin{equation}
 \rho(r)
  = \rho_{0}
    \left(\frac{r}{r_{0}}\right)^{-1}\left(1+\frac{r}{r_{0}}\right)^{-2},
\end{equation}
was realized for their dark-matter halo-formation simulations from a wide variety of initial conditions (Navarro et al.\ 1997, hereafter NFW).
In this case, the slope approaches to $-1$ toward the center.
However, Fukushige and Makino (1997) found that a shallow cusp around the center is formed because of a relatively large softening length and a small number of particles, in other words, because of the low resolution of the NFW's $N$-body simulations.
They conducted a high-resolution simulation and found that the density cusp in the center is steeper than $\rho\propto r^{-1}$.
Their result was confirmed by a follow-up work by Moore et al.\ (1998b).
Thus, it is now widely accepted that the slope of the density profile of a dark-matter halo formed in a numerical simulations with sufficiently high resolution is around $-1.5$.

However, our simulations with sufficiently high resolution have shown that the slope $r^{-1}$ is realized as the result of the dynamical evolution of the cluster and its member galaxies.

%
%

\begin{figure}[t]
 \epsfxsize\columnwidth \epsfbox{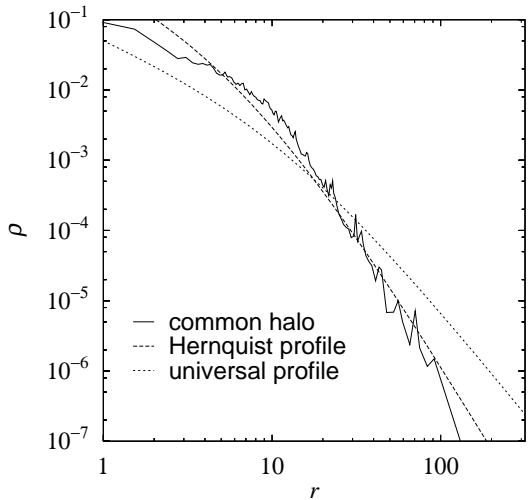}
 \caption{%
Comparison of the density profile from run A8 at $t=200$ (solid), the Hernquist profile (long dashed), and the NFW ``universal'' profile (short dashed).}
\end{figure}

Furthermore, our results show that the region expressed by $\rho\propto r^{-1}$ extends almost to the half-mass radius of the cluster.
This is quite different from NFW's ``universal'' profile, which is significantly steeper, except for in the very central region.
Figure~4 compares our results and NFW's ``universal profile''.
We also plotted the Hernquist profile.
The profile obtained in our simulation shows a sharper transition from the outer halo to the inner cusp.

At present, we do not understand the formation mechanism of the $r^{-1}$ cusp.
We are currently investigating the effect of changing initial cluster and galaxy models.

  \subsubsection{Effect of two-body relaxation}

As has been long known, two-body relaxation cause the evaporation of star clusters in the tidal field of the parent galaxy (see, e.g., Spitzer 1987).
This two-body relaxation also works in numerical simulations, and might affect the evolution of individual galaxies (Moore et al.\ 1996).

In order to see whether or not the number of particles in our numerical simulation is sufficiently large, we carried out several runs with different numbers of particles in individual galaxies (runs D1--D3).
Figure~5 shows the results.

%
%

\begin{figure}[t]
 \epsfxsize\columnwidth \epsfbox{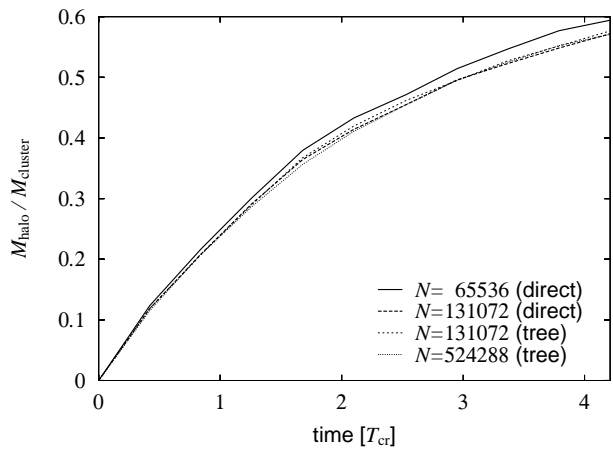}
 \caption{%
Fractional mass of the common halo $M_{\rm halo}/M_{\rm cluster}$ plotted against time in unit of the cluster crossing time.
The solid, long-dashed, short-dashed, and dotted curves denote to run D1 ($N=65536$, direct), D2 ($N=131072$, direct), D2' ($N=131072$, tree), and D3 ($N=524288$, tree), respectively.}
\end{figure}

The growth rate of the common halo for a run with $N=65536$ (D1) seems to be slightly higher than that for others.
However, the difference is small.
Furthermore, two runs with $N=131072$ (one with direct summation and the other with the tree algorithm) result in almost the same growth rate.

Here, we have shown that the growth rate of the halo, which is known to be rather sensitive to the numerical relaxation effect, is not affected by numerical relaxation in our simulations.
In addition, as shown in figure~2, the density profiles of clusters obtained by runs with different numbers of particles are practically indistinguishable.

From these results, we can safely conclude that the number of particles used in our simulations is large enough to suppress the numerical relaxation of galaxies, and that our results are reliable.
It is also confirmed that the approximation made in the tree algorithm does not affect the result.

  \subsubsection{Evolution of common halos}

We investigated how the growth rate of the common halo depends on the ratio of the size of the cluster and that of galaxies.
As described in section 2, we changed the size of the cluster as shown in table~1.

In figure~6, the mass fraction of the common halo at $t=50$ is plotted against the initial binding energy $E_{0}$ of the cluster, which is inversely proportional to the size of a cluster.
The cross symbols correspond to runs A, and the open squares are those of runs B.

This figure shows that growth of the common halo is faster for more compact clusters, if we measure the time in unit of the crossing time of the galaxies.
This is rather natural, since in a compact cluster galaxies interact more frequently.

%
%

\begin{figure}[t]
 \epsfxsize\columnwidth \epsfbox{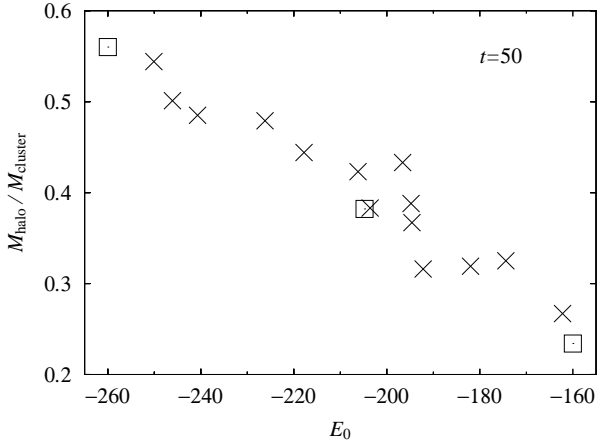}
 \caption{%
Fractional mass of the common halo, $M_{\rm halo}/M_{\rm cluster}$, at $t=50$ plotted against the energy of the cluster ($E_{0}$), for runs A (crosses) and runs B (open squares).}
\end{figure}

%
%

\begin{figure}[t]
 \epsfxsize\columnwidth \epsfbox{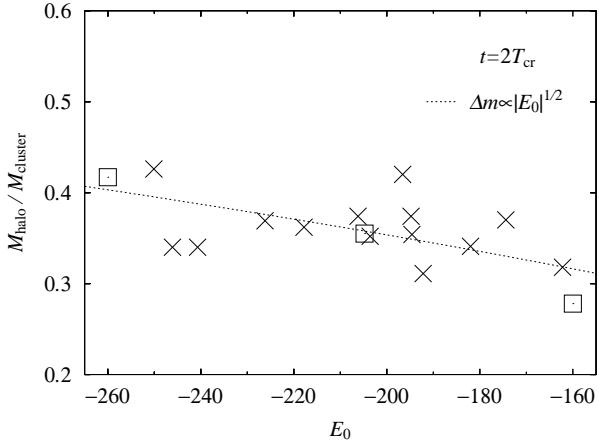}
 \caption{%
Same as figure~6, but for $t=2\,T_{\rm cr}$ instead of $t=50$.
The dotted curve indicates the theoretical prediction (see subsection 5.1).}
\end{figure}

Figure~7 is the same as figure~6, but now the mass is measured at $t=2\,T_{\rm cr}$, where $T_{\rm cr}$ is the crossing time of the cluster.
The crossing time is smaller for a more compact cluster.
Thus, the dependence of the mass on the binding energy becomes weaker than that in figure~6.

In runs A, the size of the cluster is changed due to the small-number statistics.
Thus, the cluster models, themselves, are different, and there might be some systematic effect which affects the growth rate.
In order to see the pure effect of the size of the cluster, we made additional runs (runs B), where the cluster models were the same, but scaled to different radii.
As we can see in figures~6 and 7, the results for runs B and those for runs A agree very well.

%
%

\begin{figure}[t]
 \epsfxsize\columnwidth \epsfbox{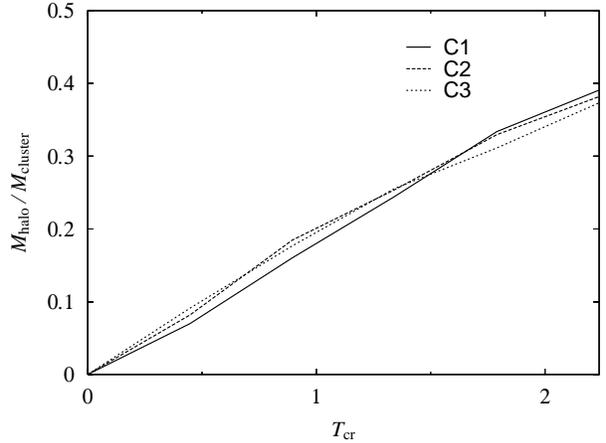}
 \caption{Same as figure~5, but for runs C1--C3.}
\end{figure}

In addition, in figure~8 we show the result of runs C1--C3, which are different models scaled so as to have exactly the same size.
Here, the growth rates are similar.
Thus, we can conclude that the size of the cluster determines the growth rate.

Note that these results are rather counter-intuitive.
If we take the standard $n\sigma v$ argument, the relation between $M_{\rm halo}$ and $E_{0}$ would be $M_{\rm halo}\propto |E_{0}|^{4}$ in figure~6 and $M_{\rm halo}\propto |E_{0}|^{2}$ in figure~7.
However, the actual power in figure~6 is around two and the dependence is very weak in figure~7.
We return to this problem in subsection 5.1.

%
%

\begin{figure}[t]
 \epsfxsize\columnwidth \epsfbox{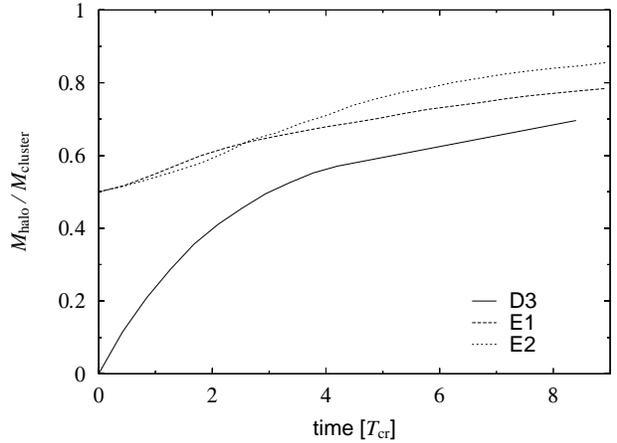}
 \caption{Same as figure~5, but for runs D3, E1, and E2.}
\end{figure}

Figure~9 shows the result of runs E, as well as that of the run D3.
From this figure, we can see that the growth rate of common halos in runs E is very similar to the part of that in the run D3 after about 4 $T_{\rm cr}$ (or, $M_{\rm halo}/M_{\rm cluster}\ge 0.5$).
We can say that the growth rate of a common halo is determined by the mass of the common halo.

 \subsection{Mass Evolution of Galaxies}

%
%

\begin{figure}[t]
 (a)\\\epsfxsize\columnwidth \epsfbox{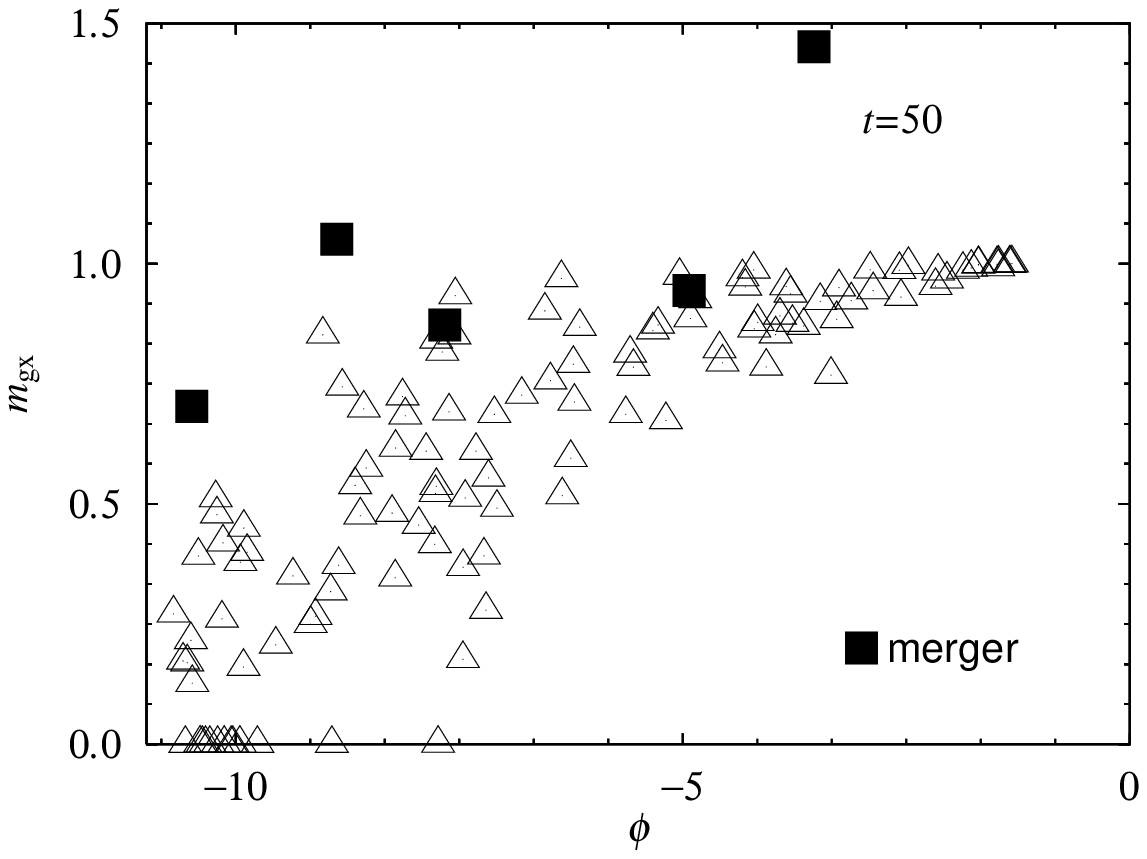}\\
 (b)\\\epsfxsize\columnwidth \epsfbox{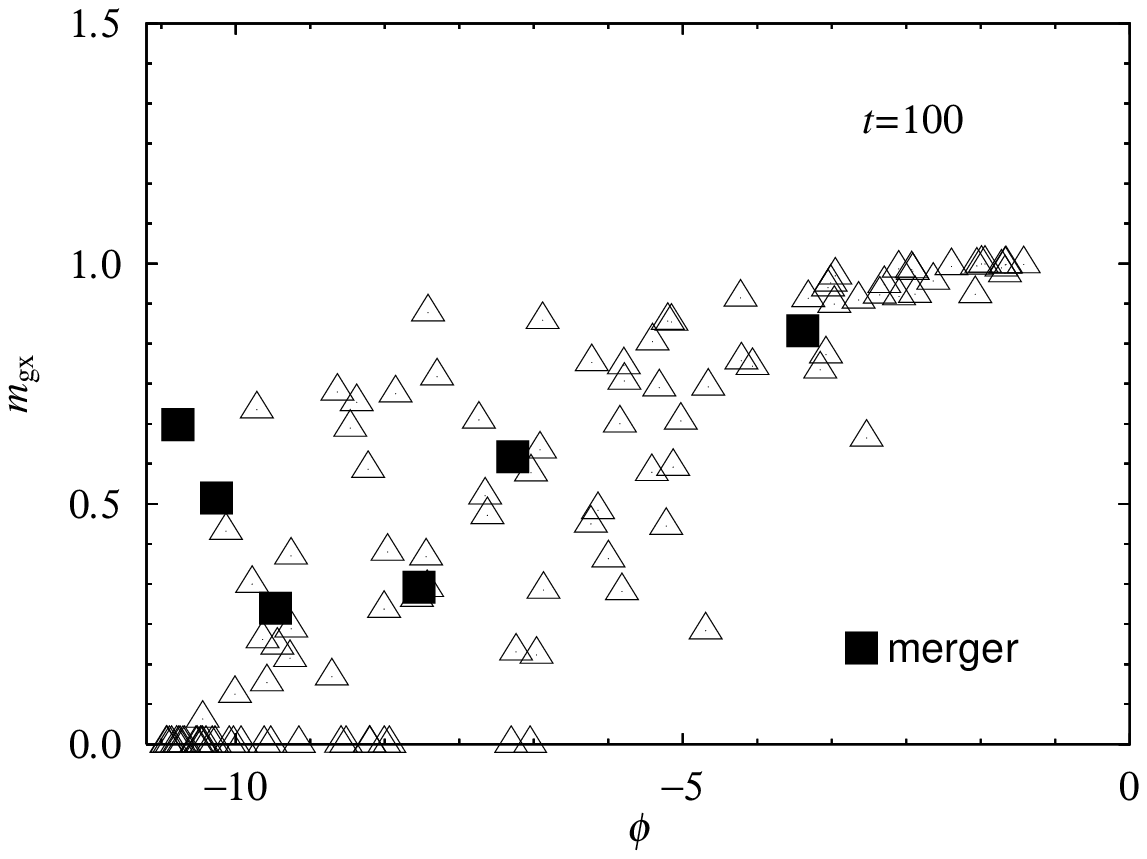}
 \caption{%
Mass of individual galaxies plotted against the depth of the cluster potential at their positions for run D3, (a) $t=50$ and (b) $t=100$.
The triangles and filled squares denote non-mergers and merger remnants, respectively.}
\end{figure}

Figure~10 shows the relation between the potential of the cluster at the positions of galaxies and their masses.
Galaxies in a deep potential well (i.e., in the central region) are smaller than those in the outer region.
This could be because the galaxy--galaxy interactions are more frequent at the center, but the stronger tidal field of the cluster as a whole might also be responsible.

In order to separate the effect of the cluster tidal field and the galaxy--galaxy encounters, we performed additional simulations in which one galaxy orbits in a fixed potential, which is smooth and has the same mass and radius as the initial cluster potential field.

%
%

\begin{figure}[t]
 \epsfxsize\columnwidth \epsfbox{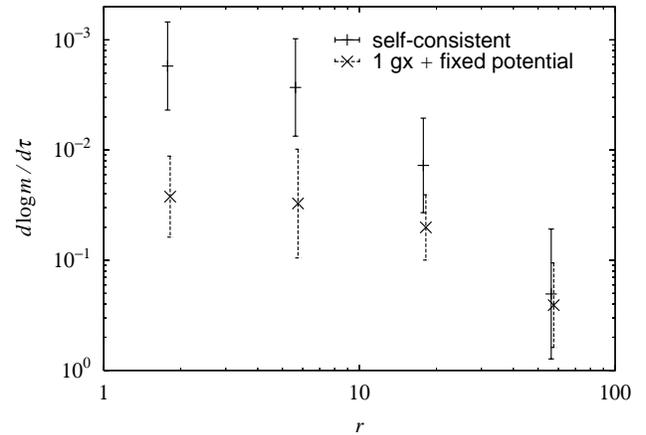}
 \caption{%
Relative mass-loss rate of the galaxies, $d\log m /d\tau$, as a function of the distance from the center.
The solid marks show the result of a self-consistent run (D3) and dashed marks show the result of the run in which one galaxy orbits in a smooth fixed cluster potential.
The error bars represents 1$\sigma$.}
\end{figure}

Figure~11 shows the relation between the distance from the center of the cluster to a galaxy and its fractional mass loss per one cluster crossing time ($1\,T_{\rm cr})$.

From this figure, it is clear that the galaxies in the self-consistent cluster models lose mass much more quickly than do those galaxies in the fixed potential.
Therefore, we can conclude that the galaxy--galaxy interaction drives the mass loss from the galaxies, and that the cluster tidal field has only a secondary effect, except possibly in the very outermost region of the cluster.
Note that our model cluster has a relatively small number of galaxies, and therefore the tidal field is actually stronger than that in typical rich clusters.
Therefore, in real clusters of galaxies, the galaxy--galaxy interaction should also be a dominant contributor to the mass loss from galaxies.

%
%

\begin{figure}[t]
 (a)\\\epsfxsize\columnwidth \epsfbox{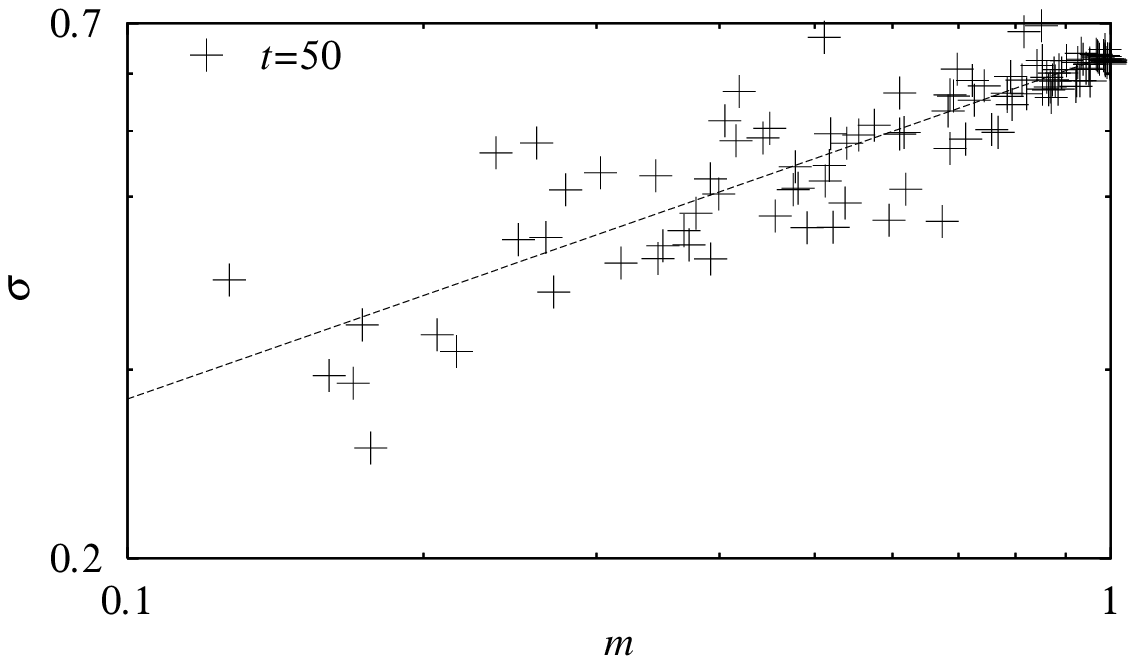}\\
 (b)\\\epsfxsize\columnwidth \epsfbox{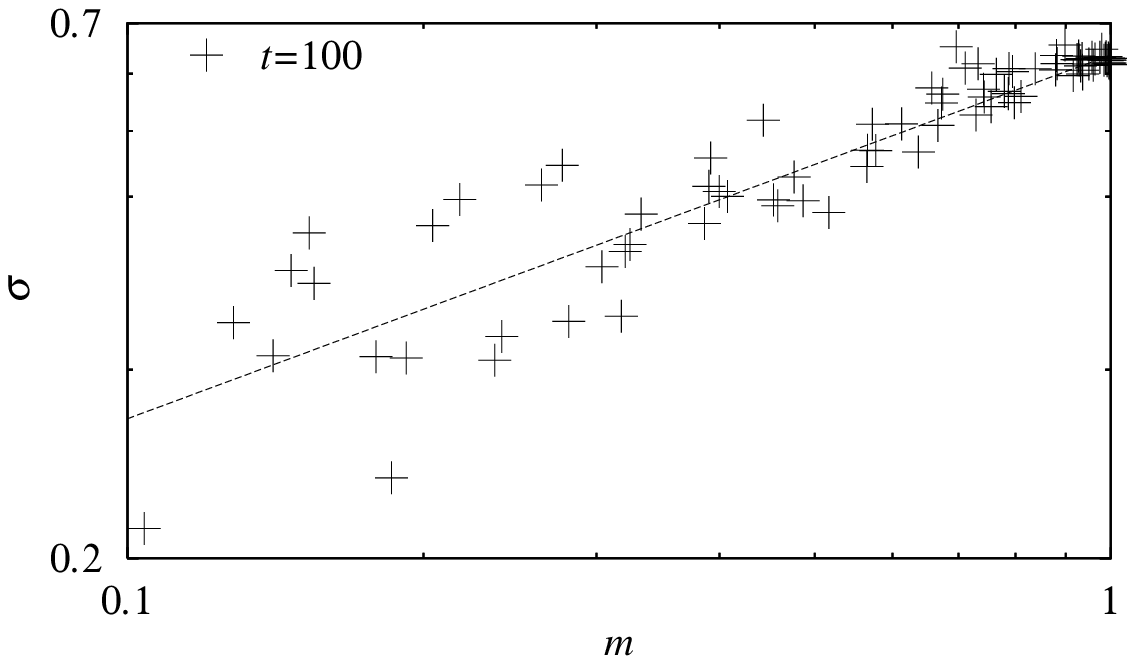}
 \caption{%
Velocity dispersions of galaxies in a cluster plotted against their masses at (a) $t=50$ and (b) $t=100$ for run D3.
The dashed lines are the best-fit line by least-squares fitting.}
\end{figure}

Figure~12 shows the relation between the masses of galaxies and their velocity dispersions.
At $t=0$, all galaxies are at the point $(1, \sqrt{0.5})$.
We can see that galaxies evolve along the line $\sigma\sim m^{1/3}$.

Funato and Makino (1999) performed a systematic study of the effect of the galaxy--galaxy encounters.
They found that the galaxies evolve along the line $\sigma\sim m^{1/4}$ if they have an extended halo of the profile $\rho\propto r^{-4}$.
If the slope of the halo is steeper, the power index of the $m$--$\sigma$ relation approaches unity.

In the present simulations, the initial galaxy model is a Plummer model.
Through interactions, an extended $r^{-4}$ halo develops, which is then truncated by the tidal field of the cluster.
Thus, we would expect the power index of the $m$--$\sigma$ relation to be somewhat larger than $1/4$, which is consistent with our numerical result.
The results shown in figure~10 and figure~12 are also consistent with the result of FME.
Therefore, we can conclude that these results are universal ones.

Figure~10 indicates that the galaxies in the central region become less massive than galaxies in the outer region.
Note that this does not necessarily imply that galaxies in the central region are fainter than those in the outer region, because the mass lost would almost entirely come from the dark halos of galaxies.
Luminous matter is more concentrated than dark matter in field galaxies.
It is reasonable to assume that cluster galaxies would also have had such an extended dark halo initially.
In this case, the luminosity of a galaxy is practically unchanged until a large fraction of its total mass is stripped (Moore et al.\ 1998a).
The morphology of galaxies is probably more sensitive to galaxy--galaxy encounters than the luminosity.
Galaxies which have lost most of the dark halo through encounters would most likely evolve to S0 or relatively faint E galaxies (Moore et al.\ 1998a).
Thus, our result naturally explains the fact that E and S0 galaxies are more frequent in the high-density region of the cluster (Dressler 1980) and the fact that young (high-$z$) clusters contain less E and S0 galaxies than low-$z$ galaxies (Dressler et al.\ 1994).

It is difficult to determine the mass of dark halos of individual galaxies from observations.
However, recently a very interesting result was reported by Natarajan (1999, also see Natarajan et al.\ 1998).
They obtained constraints on the masses of cluster galaxies from observing the gravitational lensing in clusters.
They found that the mass of galaxies in clusters increases dramatically as the
redshift increases from $z=0$ to $z=0.58$.
Their estimate is in good agreement with our result.
In our simulations, the crossing time $T_{\rm cr}$ of the cluster corresponds to 2.5 Gyr, which is roughly $\Delta z=0.2$ (assuming $h=0.65$).
Thus, $z=0.5$ corresponds to a cluster which is younger by 2--3 crossing times.

\section{Summary and Discussion}

We have followed the evolution of clusters of galaxies using self-consistent $N$-body simulations with sufficient resolution high enough to allow us to follow the evolution of individual galaxies in clusters.
We found that galaxies lose their mass rather quickly through mutual encounters, in particular in the central region.
Those galaxies which lose a large fraction of the initial mass would look like E or S0 galaxies, but they would still retain most of the luminous matter.
The matter lost from galaxies forms a common halo.

We found that the density profile of the cluster has an $r^{-1}$ cusp at the center.
The density distribution is not well fitted by a Hernquist profile, and shows a much sharper transition from the outer halo to the inner cusp.
This is a rather striking result, since initially the cluster doesn't have any cusp at all.

In the following, we first discuss the evolution timescale of the halo and its dependence on the structure of the cluster.
We then discuss the formation mechanism of the $r^{-1}$ cusp and its possible implication to observations.

 \subsection{Evolution Timescale of Common Halos}

We found that mass stripping from galaxies to the intracluster space is mainly due to the interactions between galaxies.
Their elementary process is an encounter of two galaxies, i.e., a galaxy--galaxy interaction.
Here, we try to obtain a quantitative understanding of the dependence of the mass-loss rate of individual galaxies (i.e., the growth timescale of the common halo) to the global parameters of the clusters and its member galaxies.

Funato and Makino (1999) showed, both numerically and analytically, that the mass-loss rate of a galaxy through a galaxy--galaxy interaction can be approximated as
\begin{equation}
\Delta m \propto
 m(r_{\rm h}/p)^{2} (\sigma_{\rm gx}/V)^{3},
\end{equation}
where $p$ and $V$ are the impact parameter and the relative velocity of the encounter for the range of $r_{\rm h} \le p \le p_{\rm max}$.
Here, $p_{\rm max}$ is the maximum value of the impact parameter with non-negligible mass loss.
As will be shown, the choice of $p_{\rm max}$ has only a small effect on the mass-loss rate.
For encounters with an impact parameter smaller than $r_{\rm h}$, $\Delta m$ is roughly constant.
The mass loss per unit time per galaxy can be estimated by integrating equation (7) over all possible encounters
\begin{equation}
\frac{dm}{dt} \propto
 m \int_{0}^{\infty} V^{2}\,dV
    \int_{0}^{p_{\rm max}} p\,dp\, \Delta m\,n\,Vf(V).
\end{equation}
Here, $n$ is the number density of galaxies and $f(V)$ is the distribution function of the relative velocity of galaxies.
We assume here that galaxies are distributed uniformly in the space and velocity distribution is isotropic.
Furthermore, we assume that the velocity distribution function of galaxies in the cluster is Maxwellian with velocity dispersion $V_{\rm c}$.
From these assumptions, it is straightforward to derive
\begin{equation}
\frac{dm}{dt} \propto
 \frac{mn\sigma_{\rm gx}^{3}r_{\rm h}^{2}}{V_{\rm c}^{2}}
 \log (p_{\rm max}/r_{\rm h}).
\end{equation}
The value of $p_{\rm max}/r_{\rm h}$ would be 10--300 for reasonable values of the cluster parameter, and in that range we can neglect the dependence on $p_{\rm max}$.
From the virial theorem we have $\sigma_{\rm gx}^{2}\sim m/r_{\rm h}$.
Therefore, we have
\begin{equation}
\frac{dm}{dt} \propto
 m^{5/2}n r_{\rm h}^{1/2}{V_{\rm c}^{-2}}.
\end{equation}
From equation (10), we can see that when we fix the cluster model, the mass-loss rate of individual galaxies is proportional to $r_{\rm h}^{1/2}$.
When we fix the galaxy model and change the size of the cluster, the mass-loss rate per cluster crossing time would be proportional to $|E_{0}|^{1/2}$.
The dashed line in figure~7 shows this theoretical estimate.
The agreement of the theory and numerical results is quite good.

Equation (9) clearly demonstrates why the dependence of the mass-loss rate on the size of the galaxy is weak.
When we make the size of the galaxies bigger, the geometrical effect of the $r_{\rm h}^{2}$ term increases the mass-loss rate.
This increase is, however, nearly canceled out by the increase of the encounter velocity in unit of the internal velocity dispersion of the galaxies.
Thus, the net increase in the mass-loss rate is rather modest.

Equation (10) shows that the dependence of the mass-loss rate on the mass of galaxies, itself, is rather strong.
This explains why the mass loss slows down rather quickly.

 \subsection{Properties of the Common Halo of the Cluster of Galaxies}

As found in previous simulation studies (FME), cluster-wide common halos are formed in the present simulations of clusters which initially have no common halos.
The density profile appears to be {\it universal}, in the sense that all models show the central cusp with $\rho\propto r^{-1}$.

Note, however, that the density profile obtained in our calculation is not fitted well by either the ``universal'' profile (NFW) or the Hernquist profile, though both have the $r^{-1}$ cusp at the center.
One might ask if our simulations are credible or not.

As far as the numerical accuracy is concerned, we believe that our results are okay.
We changed the number of particles from 64K to 512K and obtained essentially the same result (see figure~2).
Moreover, the local thermal relaxation time is orders of magnitude longer than the time span covered by the simulation, well into the cusp region.

On the other hand, we certainly need to explore a wider range of initial conditions.
Our initial model is a cluster made of identical galaxies, initially in virial equilibrium with no net rotation.
Real clusters are not made of identical galaxies, are initially out of dynamical equilibrium, and may have a net rotation due to tidal torque.
These differences affect the result (Garijo et al.\ 1997); a systematic survey is currently underway.

Though unexpected, theoretically it is not surprising that a central cusp develops through the dynamical evolution of a cluster.
If we regard individual galaxies as particles, the two-body relaxation time of the cluster is rather short, on the order of the crossing time of the cluster, itself.
In this case, the whole system would evolve through gravothermal catastrophe, and a nearly isothermal density cusp ($\rho\propto r^{-2.23}$) would develop (Cohn 1980).

However, the fact that each galaxy is a self-gravitating system makes the evolution of a cluster more complex.
First, galaxies lose mass and kinetic energy through close encounters.
Thus, close encounters work as a kind of dissipation, which may accelerate the collapse of the central region.
Secondly, through the dynamical friction between the common halo and the individual galaxies, galaxies would concentrate to the central region.
Galaxies behave as ``massive'' particles, while halo particles behave as light particles.
A Fokker--Planck simulation of two-component star clusters (see, e.g., Inagaki, Wiyanto 1984) have demonstrated that the central cusp of the mass distribution of light particles becomes shallower than isothermal, which qualitatively agrees with our present result.

Our result indicates that the cusp develops in a rather short timescale, in a cluster which initially has no cusp.
Simulations of cluster formation under CDM cosmology predicted steeper cusps.
Therefore, one could argue that our result is irrelevant, since in any case real clusters would have cusps right from their formation.
However, many of X-ray luminosity profiles of clusters of galaxies are consistent with the traditional $\beta$-model, which has a relatively large core with a flat density profile.
Steep cusps predicted by cosmological simulations do not fit well these ``$\beta$-model'' clusters.

As shown by Makino and Asano (1999), a $\beta$-model and the dark matter distribution of the form
\begin{equation}
 \rho\propto \frac{1}{r(1+r^{2})},
\end{equation}
have X-ray luminosity profiles which are indistinguishable from the present observations.
This profile and the profile obtained by our numerical simulations are in good agreement.
Thus, it is quite possible that these X-ray clusters actually have central cusps similar to what was obtained in the present simulations.

Several clusters are claimed to be not well fitted, at least by a single $\beta$-model, unless we accept a rather strange temperature distribution.
For these clusters, a generalized NFW profile (Makino et al.\ 1998; Suto et al.\ 1998; Makino, Asano 1998) with a central cusp steeper than $-1$ seems to give good fits.
These clusters probably have steeper cusps.

Thus, there may be a two different types of clusters of galaxies: one with steep cusps similar to the NFW profile, or even steeper ones, and the other with shallow cusps, as obtained in our simulations.
What makes the difference between these two types is not clear at this point.
This is partly because the available observations do not have sufficiently high resolutions to discriminate between models.
Observations with new X-ray telescopes with higher spatial and energy resolution, such as ASTRO-E and Chandra, will tell us the real structure of
the clusters.

\vspace{1pc}

This work was supported in part by Research for the Future Program of the Japan Society for the Promotion of Science, JSPS-RFTP 97P01102.

%
%

\section*{References}

\re
Athanassoula E., Bosma A., Lambert J.-C., Makino J. 1998, MNRAS 293, 369
\re
Barnes J. E., Hut P. 1986, Nature 324, 446
\re
Bertschinger E. 1988, ARA\&A 36, 599
\re
Bode P. W., Berrington R. C., Cohn H. N., Lugger P. M. 1994, ApJ 433, 479
\re
Cohn H. 1980, ApJ 242, 765
\re
Dressler A. 1980, ApJ 236, 351
\re
Dressler A., Oemler A. Jr, Butcher H. R., Gunn, J. E. 1994, ApJ 430, 107
\re
Evrard A. E., Silk J., Szalay A. S. 1990, ApJ 365, 13
\newpage
\re
Fukushige T., Makino J. 1997, ApJ 477, L9
\re
Funato Y., Makino J., Ebisuzaki T. 1993, PASJ 45, 289 (FME)
\re
Funato Y., Makino J. 1999, ApJ 511, 625
\re
Garijo A., Athanassoula E., Garc\'{\i}a-G\'{o}mez C. 1997, A\&A 327, 930
\re
Heggie D. C., and Mathieu R. D. 1986, in Use of Supercomputers in Stellar Dynamics, ed S. McMillan and P. Hut (Springer, Berlin), p233
\re
Inagaki S., Wiyanto P. 1984, PASJ 36, 391
\re
Makino J. 1991, PASJ 43, 621
\re
Makino J., Taiji M., Ebisuzaki T., Sugimoto, D. 1997, ApJ 480, 432
\re
Makino N., Sasaki S., Suto Y. 1998, ApJ 497, 555
\re
Makino N., Asano K. 1999, ApJ 512, 9
\re
Moore B., Katz N., Lake, G. 1996, ApJ 457, 455
\re
Moore B., Lake G., Katz N. 1998a, ApJ 495, 139
\re
Moore B., Governato F., Quinn T., Stadel J., Lake, G. 1998b, ApJ 499, L5
\re
Natarajan P., Kneib J.-P., Smail I., Ellis R. S. 1998, ApJ 499, 600
\re
Natarajan P., Kneib J.-P., Smail I. 1999, in Proc. Gravitational Lensing: Recent Progress and Future Goals, ed T. G. Brainerd and C. S. Kochanek (Boston University, Boston), in press
\re
Navarro J. F., Frenk C. S., White S. D. M. 1996, ApJ 462, 563
\re
Navarro J. F., Frenk C. S., White S. D. M. 1997, ApJ 490, 493 (NFW)
\re
Spitzer L. Jr 1987, Dynamical Evolution of Globular Clusters (Princeton University Press, Princeton), chapter 2
\re
Suto Y., Sasaki S., Makino N. 1998, ApJ 509, 544
\label{last}

\end{document}